\documentstyle[preprint,tighten,psfig,aps]{revtex}
\begin{document}
\draft

\title{ $W$ and $Z$ Boson Production in {\mbox{$p\bar p$}}\
Collisions at {\mbox{$\sqrt{s}$ =\ 1.8\ TeV}} }

\author{
S.~Abachi,$^{12}$
B.~Abbott,$^{33}$
M.~Abolins,$^{23}$
B.S.~Acharya,$^{40}$
I.~Adam,$^{10}$
D.L.~Adams,$^{34}$
M.~Adams,$^{15}$
S.~Ahn,$^{12}$
H.~Aihara,$^{20}$
J.~Alitti,$^{36}$
% ...
\\
\vskip 0.25cm
\centerline{(D\O\ Collaboration)}
\vskip 0.25cm
}
\address{
\centerline{$^{1}$Universidad de los Andes, Bogot\'{a}, Colombia}
\centerline{$^{2}$University of Arizona, Tucson, Arizona 85721}
\centerline{$^{3}$Brookhaven National Laboratory, Upton, New York 11973}
\centerline{$^{4}$Brown University, Providence, Rhode Island 02912}
\centerline{$^{5}$University of California, Davis, California 95616}
\centerline{$^{6}$University of California, Irvine, California 92717}
\centerline{$^{7}$University of California, Riverside, California 92521}
\centerline{$^{8}$LAFEX, Centro Brasileiro de Pesquisas F{\'\i}sicas,
                  Rio de Janeiro, Brazil}
\centerline{$^{9}$CINVESTAV, Mexico City, Mexico}
\centerline{$^{10}$Columbia University, New York, New York 10027}
\centerline{$^{11}$Delhi University, Delhi, India 110007}
\centerline{$^{12}$Fermi National Accelerator Laboratory, Batavia,
                   Illinois 60510}
\centerline{$^{13}$Florida State University, Tallahassee, Florida 32306}
\centerline{$^{14}$University of Hawaii, Honolulu, Hawaii 96822}
\centerline{$^{15}$University of Illinois at Chicago, Chicago, Illinois 60607}
\centerline{$^{16}$Indiana University, Bloomington, Indiana 47405}
\centerline{$^{17}$Iowa State University, Ames, Iowa 50011}
\centerline{$^{18}$Korea University, Seoul, Korea}
\centerline{$^{19}$Kyungsung University, Pusan, Korea}
\centerline{$^{20}$Lawrence Berkeley Laboratory and University of California,
                   Berkeley, California 94720}
\centerline{$^{21}$University of Maryland, College Park, Maryland 20742}
\centerline{$^{22}$University of Michigan, Ann Arbor, Michigan 48109}
\centerline{$^{23}$Michigan State University, East Lansing, Michigan 48824}
\centerline{$^{24}$Moscow State University, Moscow, Russia}
\centerline{$^{25}$University of Nebraska, Lincoln, Nebraska 68588}
\centerline{$^{26}$New York University, New York, New York 10003}
\centerline{$^{27}$Northeastern University, Boston, Massachusetts 02115}
\centerline{$^{28}$Northern Illinois University, DeKalb, Illinois 60115}
\centerline{$^{29}$Northwestern University, Evanston, Illinois 60208}
\centerline{$^{30}$University of Notre Dame, Notre Dame, Indiana 46556}
\centerline{$^{31}$University of Panjab, Chandigarh 16-00-14, India}
\centerline{$^{32}$Institute for High Energy Physics, 142-284 Protvino, Russia}
\centerline{$^{33}$Purdue University, West Lafayette, Indiana 47907}
\centerline{$^{34}$Rice University, Houston, Texas 77251}
\centerline{$^{35}$University of Rochester, Rochester, New York 14627}
\centerline{$^{36}$CEA, DAPNIA/Service de Physique des Particules, CE-SACLAY,
                   France}
\centerline{$^{37}$Seoul National University, Seoul, Korea}
\centerline{$^{38}$State University of New York, Stony Brook, New York 11794}
\centerline{$^{39}$SSC Laboratory, Dallas, Texas 75237}
\centerline{$^{40}$Tata Institute of Fundamental Research,
                   Colaba, Bombay 400005, India}
\centerline{$^{41}$University of Texas, Arlington, Texas 76019}
\centerline{$^{42}$Texas A\&M University, College Station, Texas 77843}
}

\date{\today}

\maketitle

\vskip -0.7cm
\begin{abstract}
\vskip -1.0 cm
The inclusive cross sections times leptonic branching ratios
for $W$ and $Z$ boson production in {\mbox{$p\bar p$}}\ collisions at
{\mbox{$\sqrt{s}$ =\ 1.8\ TeV}}\
were measured using the D\O\ detector at the Fermilab Tevatron collider:
\begin{eqnarray}
\sigma_W \cdot B({\mbox{$ W\rightarrow e \nu$}})\; \ &&= \
2.36 \pm 0.07  \pm 0.13 {\rm~nb} \nonumber\\
\sigma_W \cdot B({\mbox{$ W\rightarrow \mu \nu$}})\; \ &&= \
2.09 \pm 0.23 \pm 0.11 {\rm~nb} \nonumber\\
\sigma_Z \cdot B({\mbox{$ Z\rightarrow {e^+e^-}$}}) \; \ &&= \
0.218 \pm 0.011 \pm 0.012 {\rm~nb} \nonumber\\
\sigma_Z \cdot B({\mbox{$ Z\rightarrow {\mu^+\mu^-}$}})\; \ &&= \
0.178 \pm 0.030 \pm 0.009 {\rm~nb} \nonumber
\end{eqnarray}
The first error is the combined
statistical and systematic uncertainty, and the second reflects the
uncertainty in the luminosity.
For the combined electron
and muon analyses we find
${\sigma_W \cdot B(W\rightarrow l\nu)}/
{\sigma_Z \cdot B(Z\rightarrow l^+l^-)}=
10.90\pm 0.49$.
Assuming Standard Model couplings, this result is used to determine
the width of the $W$ boson, $\Gamma(W) = 2.044 \pm 0.093 \; {\rm GeV}$.
\end{abstract}

%\vskip 2.0cm
\pacs{PACS numbers: 13.38.-b, 13.85.Qk, 14.65.Ha, 14.70.Fm, 14.70.Hp}

The measurement of the production cross sections times leptonic branching
ratios
($\sigma \cdot B$) for $W$ and $Z$ bosons allows a
determination of the width of the $W$ boson and a comparison of $W$ and $Z$
boson production with QCD predictions.
The total width of the $Z$ boson is known to
a precision of $0.3\%$~\cite{PDG}, which
places strong constraints on the existence of new particles produced
in neutral weak decays.  Our knowledge of the
total width of the $W$ boson is an order of
magnitude less precise, and the
corresponding limits on charged weak decays are much less stringent.
It is therefore important to improve the measurement of the $W$ boson width
as a means of searching for unexpected $W$ boson decay modes.

We determine the leptonic branching ratio of the $W$ boson,
$B(W\rightarrow l\nu)$,
from the ratio of the measured $W$ and $Z$ boson $\sigma \cdot B$ values
\begin{equation}
R \equiv {{\sigma_W \cdot B(W\rightarrow l\nu)}\over
{\sigma_Z \cdot B(Z\rightarrow ll)}},
\label{Reqn}
\end{equation}
where $l=e$ or $\mu$,
$\sigma_W$ and $\sigma_Z$ are the inclusive
cross sections for $W$ and $Z$ boson production in {\mbox{$p\bar p$}}\
collisions,
and $B(Z\rightarrow ll)$ is the leptonic branching ratio of the $Z$ boson.
We extract $B(W\rightarrow l\nu)$ from the
above ratio using a theoretical calculation of $\sigma_W/ \sigma_Z$
and the precise measurement of $B(Z\rightarrow ll)$
from LEP.  We then combine
$B(W\rightarrow l\nu)$ with a theoretical calculation of
the $W$ boson leptonic partial width, ${\Gamma(W\rightarrow l\nu)}$,
to obtain the $W$ boson total width, $\Gamma(W)$.
Previous measurements of ${\Gamma(W)}$ have been published by
UA1~\cite{UA1}, UA2~\cite{UA2} and CDF~\cite{CDF1,CDF2}.

In this letter, we report a new measurement of $\sigma \cdot B$
and determination of ${\Gamma(W)}$
using data collected with the D\O\ detector~\cite{D0NIM}
in 1992-93 at the Fermilab Tevatron
{\mbox{$p\bar p$}}\ collider at {\mbox{$\sqrt{s}$ =\ 1.8\ TeV}}. The four decay
channels included in this analysis are {\mbox{$ W\rightarrow e \nu, \mu \nu$}}
and {\mbox{$ Z\rightarrow {e^+e^-},{\mu^+\mu^-}$}}.

Electrons were detected in hermetic,
uranium liquid-argon calorimeters~\cite{EMNIM,TBNIM},
with an energy resolution of
about $15\%/\sqrt{E (\mbox{GeV})}$.
The calorimeters
have a transverse granularity of $\Delta\eta \times
\Delta\phi = 0.1 \times 0.1$, where $\eta$ is the pseudorapidity
and $\phi$ is the azimuthal angle.

For the {\mbox{$ W\rightarrow e \nu$}}\ and {\mbox{$ Z\rightarrow {e^+e^-}$}}\
analyses we accepted electrons with $|\eta|<1.1$ or $1.5 < |\eta| < 2.5$.
The $W$ and $Z$ boson analyses both used the same trigger which required
a single electron with transverse
energy ($E_T$) greater than 20 GeV.
Kinematic selections were made in the offline analysis
requiring that $Z$ boson candidates have two
electrons, each with
$E_T > 25$ GeV, and that
$W$ boson candidates have one electron with $E_T > 25$ GeV and
missing transverse energy ({\mbox{$\not\!\!E_T$}}) greater than 25 GeV.

\begin{figure}
\centerline{\psfig{bbllx=15pt,bblly=219pt,bburx=584pt,bbury=620pt,figure=
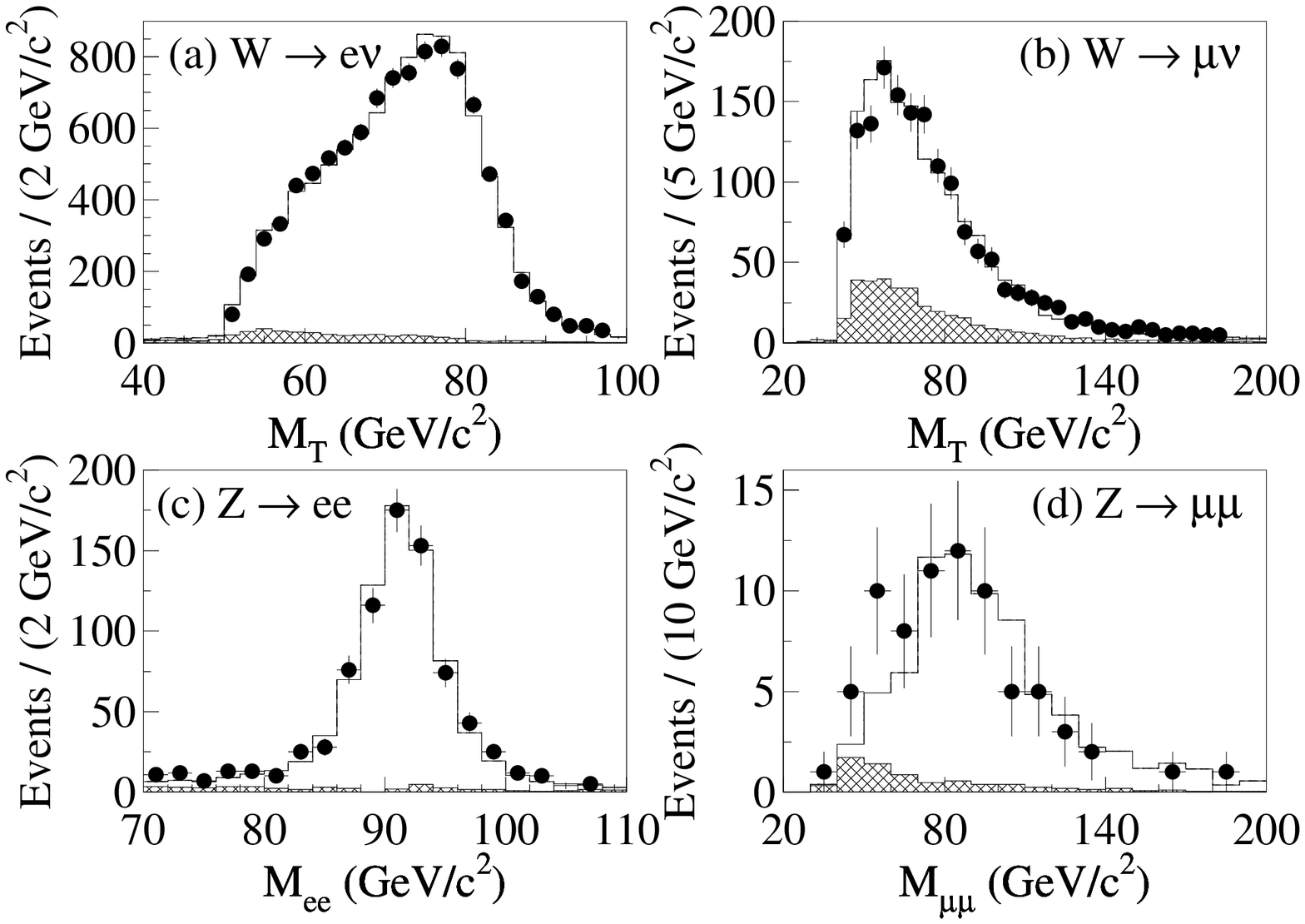,height=11.3cm,width=16.0cm}}
\vskip 0.3cm
\caption{Transverse mass and invariant mass distributions for the
indicated channels. The points are the data.
The shaded areas represent the estimated backgrounds, and the solid
lines correspond to the sums of the expected signals (from Monte Carlo)
and the estimated backgrounds.}
\label{spectra}
\end{figure}

The offline electron identification requirements
consisted of three criteria for a ``loose'' electron:
{\it i}) the electron had to deposit at least 95\% of its energy in the 21
radiation length electromagnetic calorimeter,
{\it ii}) the transverse and longitudinal shower shapes had to be
consistent with
those expected for an electron (based on test beam measurements), and
{\it iii}) the electron had to be isolated with $\it{I}<$~0.1.
The isolation variable
is defined as $\it{I}$=($E_{\rm tot}$(0.4)-$E_{\rm EM}$(0.2))/$E_{\rm
EM}$(0.2),
where $E_{\rm tot}$(0.4)
is the total calorimeter energy inside a cone of radius $\sqrt{\Delta\eta^2
+\Delta\phi^2} = 0.4$ and $E_{\rm EM}$(0.2) is the electromagnetic energy
inside a cone of 0.2.
For a ``tight'' electron we also required
a good match between a reconstructed track in the drift
chamber system and the shower position in the calorimeter.
For $W$ boson events we required one ``tight'' electron, while for
$Z$ boson events we required one electron to be ``tight'' and the other
to be either ``tight'' or ``loose.''
Figures \ref{spectra}(a) and \ref{spectra}(c) show the observed transverse mass
and invariant mass distributions for {\mbox{$ W\rightarrow e \nu$}}\
and {\mbox{$ Z\rightarrow {e^+e^-}$}}\ candidates passing these cuts.
For the {\mbox{$ Z\rightarrow {e^+e^-}$}}\ analysis we used the events in the
invariant mass range $75-105$~GeV/c$^2$.

The kinematic and geometric acceptance (shown in Table I) for the
{\mbox{$ W\rightarrow e \nu$}}\ and {\mbox{$ Z\rightarrow {e^+e^-}$}}\ channels
was calculated
with a Monte Carlo simulation using the measured detector resolutions to smear
generated four-momenta. The calculation used the
CTEQ2M~\cite{cteq} parton distribution functions (pdf), and a
NLO calculation~\cite{AK} of the $W$ boson transverse
momentum.
The systematic error in the acceptance includes contributions from
the uncertainty in the pdf (the spread among
CTEQ2~\cite{cteq}, MRS~\cite{mrs},
and GRV~\cite{grv} pdf),
from the uncertainty in
the $W$ boson mass~\cite{wmass}, and from systematic
uncertainties associated with modeling the detector response.
The trigger and selection efficiencies (Table I) were determined using
{\mbox{$ Z\rightarrow {e^+e^-}$}}\ events where one of the electrons satisfied
tight trigger and selection criteria and the second electron
provided an unbiased sample to measure the efficiencies.  The trigger
efficiencies were found to be $>95\%$.

Muons were detected as tracks in three layers of proportional
drift tube (PDT) chambers outside the calorimeter.
One layer of PDT chambers had four planes and was located inside an iron
toroid magnet.  The other two layers, each with three planes, were located
outside of the iron.
The muon momentum resolution in this analysis was
$\sigma(1/p) = 0.18(p-2)/p^2~\oplus~0.008$ (with $p$ in GeV/c).
A muon track was required to match a charged track in the central drift
chamber system.
We accepted muons that passed through the central
iron toroid ($|\eta|<1.0$).

The {\mbox{$ W\rightarrow \mu \nu$}}\ and
{\mbox{$ Z\rightarrow {\mu^+\mu^-}$}}\
analyses both used the same trigger which required
a single muon with transverse
momentum ($p_T$) greater than 15 GeV/c.
Cosmic ray background was reduced
by rejecting muons that also had hits or tracks
within 10 degrees in $\theta$ and 20 degrees in $\phi$
in the muon chambers on the opposite side of the interaction point.
Trigger efficiencies were measured using the subsample of
events with high $p_T$ muons that satisfied
jet or {\mbox{$\not\!\!E_T$}}\ triggers, and also using the second muon in
{\mbox{$ Z\rightarrow {\mu^+\mu^-}$}}\ events.
The trigger efficiency was about $40$\% ($70$\%) for $W$($Z$) boson events.
Kinematic cuts were made requiring muon $p_T>20$ GeV/c and
{\mbox{$\not\!\!E_T$}}$> 20$ GeV
for $W$ boson events, and $p_T > 15,20$ GeV/c for the two muons in $Z$ boson
events.

A ``loose'' muon was required to deposit sufficient energy in the calorimeter
to be consistent with the passage of a minimum ionizing particle and
to traverse a minimum field integral of 2.0 T$\cdot$m.  This
latter requirement restricts the muon analysis to a region of the detector with
$\ge 13$ interaction lengths of material, so that hadronic
punchthrough is negligible.
A ``tight'' muon had five additional requirements: {\it i}) a stringent
track match with a track in the central detector,
{\it ii}) a good quality global fit with the vertex and
a central detector track,
{\it iii}) a muon time of origin within 100 ns of the beam crossing,
{\it iv}) energy in the calorimeter consistent with single muon ionization
within a cone of radius $\sqrt{\Delta\eta^2+\Delta\phi^2}=0.2$
and with less than 6 GeV of additional energy in a cone of 0.6, and
{\it v}) a good muon impact parameter.

For $Z$ boson events, we required at least one muon to be ``tight'' and
the other
to be either ``tight'' or ``loose.''
For $W$ boson events, we required at least one ``tight'' muon (after
$Z$ boson candidates were removed).
Figures \ref{spectra}(b) and \ref{spectra}(d) show the observed transverse mass
and invariant mass distributions for {\mbox{$ W\rightarrow \mu \nu$}}\ and
{\mbox{$ Z\rightarrow {\mu^+\mu^-}$}}\ candidates passing our criteria.
The kinematic and geometric acceptances (Table I) were calculated with a
full detector Monte Carlo simulation.
The selection efficiencies (Table I) were
determined with {\mbox{$ Z\rightarrow {\mu^+\mu^-}$}}\ events
using the same method that was used for electrons.

The background estimates (Table I) due to
{\mbox{$ Z\rightarrow {e^+e^-}$}}\ or
{\mbox{$ Z\rightarrow {\mu^+\mu^-}$}}\ (where one of
the electrons or muons was lost) and $W\rightarrow\tau\nu$
or $Z \rightarrow \tau^+ \tau^-$ (where
$\tau\rightarrow e \nu\nu$ or $\mu\nu\nu$) were obtained from
Monte Carlo.
The multijet background estimate for the $W\rightarrow e\nu$ sample
was derived from
the data by measuring the tail of the
{\mbox{$\not\!\!E_T$}}\ distribution
for non-isolated electrons and normalizing this at small
{\mbox{$\not\!\!E_T$}} to the
{\mbox{$\not\!\!E_T$}}\ spectrum for isolated electrons.
The multijet background in the {\mbox{$ W\rightarrow \mu \nu$}}\ and
{\mbox{$ Z\rightarrow {\mu^+\mu^-}$}}\ samples was estimated using the
distribution of the isolation variable.
The amount of multijet background in the {\mbox{$ Z\rightarrow {e^+e^-}$}}\
sample was estimated by performing a
fit to the data using the predicted $Z$ boson mass distribution and the
experimentally
determined shape of the multijet background from dijet and direct photon
events.
The cosmic ray and random hit backgrounds to {\mbox{$ W\rightarrow \mu \nu$}}\
and {\mbox{$ Z\rightarrow {\mu^+\mu^-}$}}\
were estimated from the distributions of muon time of origin relative to
beam crossing.

The luminosity (Table~I) was calculated by measuring the rate
for {\mbox{$p\bar p$}}\
non-diffractive inelastic collisions using two
hodoscopes of scintillation counters~\cite{D0NIM}
mounted close to the
beam on the front surfaces of the end calorimeters.
The normalization for the
luminosity measurement and the $5.4\%$ systematic error
in the luminosity, which has been estimated from the uncertainty in the
{\mbox{$p\bar p$}}\
inelastic cross section and uncertainties in the acceptance and efficiency of
the counters, are described in Ref.\ \cite{lum}.

\begin{figure}
\centerline{\psfig{bbllx=23pt,bblly=286pt,bburx=575pt,bbury=568pt,figure=
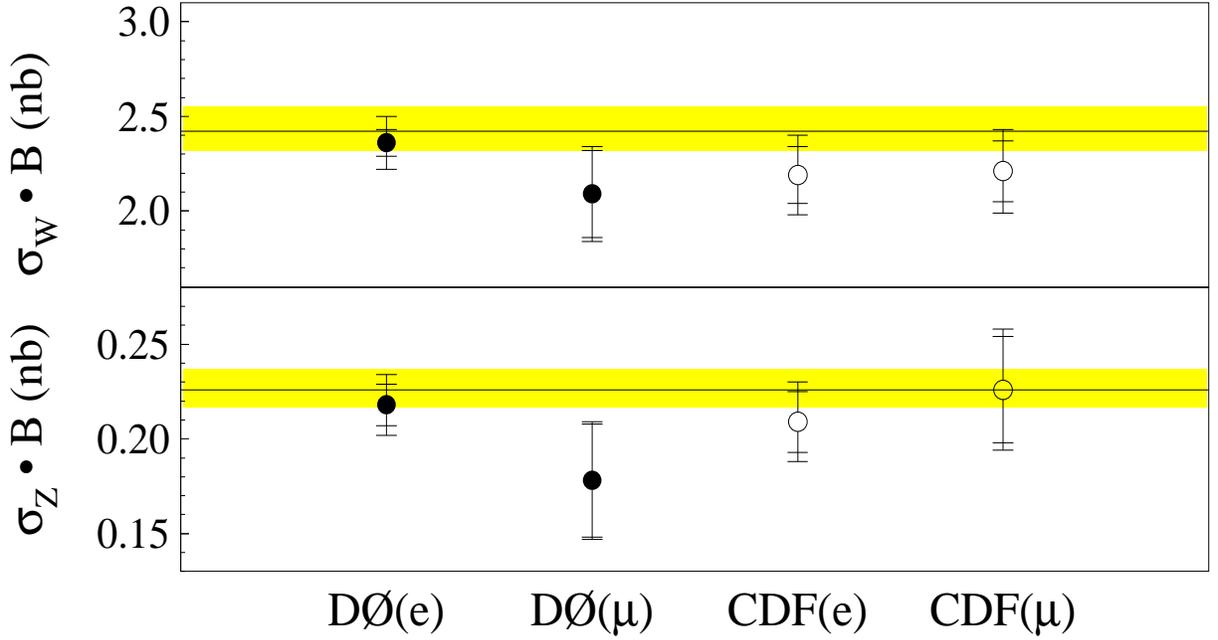,height=8.5cm,width=16.0cm}}
\vskip 0.3cm
\caption{$\sigma\cdot B$ for inclusive $W$ and $Z$ boson production.
The inner error bar is the combined statistical and
systematic uncertainty and the outer error bar includes the luminosity
uncertainty. The solid line and shaded band are the theoretical prediction
described in the text.}
\label{wzx}
\end{figure}

We calculated $\sigma\cdot B$ by subtracting the background from the number
of observed events ($N_{\rm obs}$), and dividing the result by the acceptance,
efficiency, and luminosity.
The results obtained
are shown in Table~I, and are plotted in Fig.\ \ref{wzx},
together with the CDF results~\cite{CDF3,CDF4} and the
theoretical ${\cal{O}}(\alpha_s^2)$ QCD prediction~\cite{xsec1,xsec2} of
$\sigma_W\cdot B(W\rightarrow l\nu) = 2.42^{+0.13}_{-0.11}$~nb and
$\sigma_Z\cdot B(Z\rightarrow ll) = 0.226^{+0.011}_{-0.009}$~nb.

Using the definition for $R$ in Eq.~(1),
we obtain for the electron, muon and combined results:
\begin{eqnarray}
R_e \ &&= \
10.82\pm{0.41(stat)}\pm{0.30(sys)},\nonumber\\
R_{\mu} \ &&= \
11.8^{+1.8}_{-1.4}(stat)\pm{1.1(sys)}, {\rm\, and} \nonumber\\
R_{e+\mu} \ &&= \
10.90 \pm 0.49(stat \oplus sys)\nonumber.
\end{eqnarray}
Many common sources of error cancel in $R$,
including the uncertainty in the luminosity and parts of the
errors in the acceptance and event selection efficiency.

We use the combined ratio $R_{e+\mu}$ and Eq.~(1) to determine
$B(W\rightarrow l\nu)$.
We use
$B(Z\rightarrow ll)= (3.367 \pm 0.006)\%$~\cite{PDG}, and
the theoretical calculation~\cite{xsec1} of ${\sigma_W/ \sigma_Z}=
3.33\pm 0.03$, where the quoted error is
due to systematic differences in the pdf choices~\cite{cteq,mrs}
(with CTEQ2ML and CTEQ2MS giving the maximum variation) and the
uncertainty in the $W$ boson mass~\cite{wmass}.
We obtain
\begin{eqnarray}
{B(W\rightarrow l\nu) = (11.02 \pm 0.50)\%.}\nonumber
\end{eqnarray}

We combine this measurement of $B(W\rightarrow l\nu)$ with a
theoretical calculation~\cite{rosner,wmass} of ${\Gamma(W\rightarrow l\nu)} =
225.2 \pm 1.5$ MeV to obtain
\begin{eqnarray}
\Gamma(W) = 2.044 \pm 0.093\; {\rm GeV.}\nonumber
\end{eqnarray}

The measurement of $\Gamma(W)$ (or $B(W\rightarrow l\nu)$) can be used to set
limits on unexpected decay modes of the $W$ boson, such as $W$ decays into
supersymmetric charginos and neutralinos~\cite{susy}, or into heavy
quarks~\cite{alvarez}.  Comparing our result for $\Gamma(W)$ with the Standard
Model prediction, $\Gamma(W) = 2.077 \pm 0.014$ GeV~\cite{rosner,wmass}, the
95\% CL upper limit on the contribution of unexpected decays to the $W$ boson
width is 164 MeV.  Combining our result for $\Gamma(W)$ with other
measurements~\cite{average} gives a weighted average of
$\Gamma(W) = 2.062 \pm 0.059$ GeV.  Comparing this weighted average with the
Standard Model value gives a 95\% CL upper limit of 109 MeV on unexpected
decays.

We thank the Fermilab Accelerator, Computing, and Research Divisions, and
the support staffs at the collaborating institutions for their contributions
to the success of this work.   We also acknowledge the support of the
U.S. Department of Energy,
the U.S. National Science Foundation,
the Commissariat \`a L'Energie Atomique in France,
the Ministry for Atomic Energy and the Ministry of Science and
Technology Policy in Russia,
CNPq in Brazil,
the Departments of Atomic Energy and Science and Education in India,
Colciencias in Colombia, CONACyT in Mexico,
the Ministry of Education, Research Foundation and KOSEF in Korea
and the A.P. Sloan Foundation.

\begin{table*}
\squeezetable
\label{all}
\caption{Production cross section times leptonic branching ratio for
$W$ and $Z$ bosons.}
\vskip 0.35 cm
\begin{tabular}{l|c|c|c|c}
\hspace{.7 cm}
Channel                             & $W\rightarrow{e}\nu$    &
                                      $Z\rightarrow{e^+e^-}$  &
                                      {\mbox{$ W\rightarrow \mu \nu$}}      &
                                      {\mbox{$ Z\rightarrow {\mu^+\mu^-}$}}  \\
\hline
\hspace{1 cm}
$N_{\rm obs}$                             & $ 10338 $         &
                                            $   775 $         &
                                            $  1665 $         &
                                            $    77 $         \\
\hline
\hspace{.3 cm}
Backgrounds(\%):                      &                           &
                                                                  &
                                                                  &
                                                                  \\
\hspace{.6 cm}
$Z\rightarrow ee, \mu\mu, \tau\tau$   & $0.6 \pm 0.1 $         &
                                         ---                   &
                                        $7.3 \pm 0.5 $         &
                                        $0.7 \pm 0.2 $         \\
\hspace{.6 cm}
$W\rightarrow \tau\nu$                & $1.8 \pm 0.1 $         &
                                         ---                   &
                                        $5.9 \pm 0.5 $         &
                                         ---                   \\
\hspace{.6 cm}
Multijet                              & $3.3 \pm 0.5 $         &
                                        $2.8 \pm 1.4 $         &
                                        $5.1 \pm 0.8 $         &
                                        $2.6 \pm 0.8 $         \\
\hspace{.6 cm}
Cosmic/Random                         &  ---                  &
                                         ---                  &
                                        $3.8 \pm 1.6 $        &
                                        $5.1 \pm 3.6 $        \\
\hspace{.6 cm}
Drell-Yan                             &  ---                  &
                                        $1.2 \pm 0.1 $        &
                                         ---                  &
                                        $1.7 \pm 0.3 $        \\
\hspace{.3 cm}
Total Bkgnd(\%)                       & $ 5.7 \pm 0.5 $         &
                                        $ 4.0 \pm 1.4 $         &
                                        $22.1 \pm 1.9 $         &
                                        $10.1 \pm 3.7 $         \\
\hline
\hspace{.4 cm}
Acceptance(\%)                      & $46.0 \pm 0.6 $       &
                                      $36.3 \pm 0.4 $       &
                                      $24.8 \pm 0.7 $       &
                                      $ 6.5 \pm 0.4 $       \\
\hline
\hspace{.5 cm}
$\epsilon_{\rm trig}*\epsilon_{\rm sel}$(\%)  & $70.4 \pm 1.7 $       &
                                              $73.6 \pm 2.4 $       &
                                              $21.9 \pm 2.6 $       &
                                              $52.7 \pm 4.9 $       \\
\hline
\hspace{.4 cm}
$\rm \int{{\cal{L}}dt}\ ({\mbox{${\rm pb}^{-1}$}})$  & $ 12.8\pm 0.7  $       &
                                        $ 12.8\pm 0.7  $       &
                                        $ 11.4\pm 0.6  $       &
                                        $ 11.4\pm 0.6  $       \\
\hline
\hspace{.6 cm}
{$\bf\sigma\cdot{\rm B}$\enskip (nb)}  &  $ 2.36   $      &
                                          $ 0.218  $      &
                                          $ 2.09   $      &
                                          $ 0.178  $      \\
$\pm(stat),(sys),(lum)$  & $\pm 0.02  \pm 0.07  \pm 0.13   $ \hspace{0.20cm} &
                           $\pm 0.008 \pm 0.008 \pm 0.012  $ \hspace{0.20cm} &
                           $\pm 0.06  \pm 0.22  \pm 0.11   $ \hspace{0.20cm} &
                           $\pm 0.022 \pm 0.021 \pm 0.009  $ \hspace{0.20cm} \\
\end{tabular}
\end{table*}


\begin{references}
\bibitem{xsec1} We calculated $\sigma_W = 22.35$~nb and $\sigma_Z = 6.708$~nb
with Ref.~\cite{HVM}, using the CTEQ2M pdf~\cite{cteq},
$M_Z=91.19$ GeV/c$^2$~\cite{PDG}, $M_W$ from Ref.~\cite{wmass}, and
$\sin^2\theta_W \equiv 1-(M_W/M_Z)^2 = 0.2259$.
\bibitem{xsec2} To obtain the predicted $\sigma\cdot B$ we used $\sigma$ from
Ref.~\cite{xsec1},
$B(W\rightarrow l\nu) = (10.84 \pm 0.02)\%$~\cite{rosner}, and
$B(Z\rightarrow ll) = (3.367 \pm 0.006)\%$~\cite{PDG}.
The errors in $\sigma \cdot B$ include a $(2.5-4.5)\%$ error due to the
effect of using
various pdf choices~\cite{cteq,mrs}
(with CTEQ2ML and CTEQ2MS giving the maximum variation),
a $3\%$ error due to using NLO pdf instead of NNLO~\cite{NZ},
a $(0.2-0.6)\%$ error due to the uncertainty in $M_W$~\cite{wmass}, and
a $(0.3-0.5)\%$ error due to
the variation of the renormalization and factorization scales~\cite{stirling}.
\bibitem{average} We recalculated $\Gamma(W)$ for the other
experiments~\cite{UA1,UA2,CDF1} by using their published $R$ values along
with the values of $B(Z\rightarrow ll)$ and
${\Gamma(W\rightarrow l\nu)}$ in this paper. For ${\sigma_W/ \sigma_Z}$ we
used our value for CDF, and
$3.26 \pm 0.09$~\cite{UA2} for the UA1 and UA2 experiments at the
lower~$\sqrt{s}$.
\end{references}
\end{document}